\documentclass[draftcls,onecolumn]{IEEEtran}

\usepackage[pdftex]{graphicx}

\ifCLASSOPTIONcompsoc
    \usepackage[caption=false, font=normalsize, labelfont=sf, textfont=sf]{subfig}
\else
\usepackage[caption=false, font=footnotesize]{subfig}
\fi

\usepackage{amssymb}
\usepackage{amsmath}
\usepackage{cite}
\usepackage{enumerate}
\usepackage{amsthm}
\usepackage{bm}
\usepackage{subfig}
\usepackage{psfrag}
\usepackage{cases}
\usepackage{mathtools}
\usepackage{float}
\usepackage{stfloats}
\usepackage{ifthen}
\usepackage[capitalize]{cleveref}
\usepackage{algorithm}
\usepackage{algorithmic}
\usepackage{booktabs}
\usepackage[usenames,dvipsnames]{color}
\usepackage{bbm}
\usepackage{lipsum}

\ifCLASSINFOpdf
\usepackage{graphicx}
\DeclareGraphicsExtensions{.pdf,.jpeg,.png,.eps}
\else
\DeclareGraphicsExtensions{.eps}
\fi
\usepackage{tikz}
\usetikzlibrary{arrows}
\usepackage{makecell}

\newcommand{\norm}[1]{\left\lVert#1\right\rVert}
\renewcommand{\(}{\left(}
\renewcommand{\)}{\right)}
\renewcommand{\[}{\left[}
\renewcommand{\]}{\right]}

\newboolean{draft}
\newcommand{\isdraft}[2]{\ifthenelse{\boolean{draft}}{#1}{#2}}

\isdraft{\usepackage{setspace}}{}                              
\isdraft{\usepackage{paralist}}{}                              

\def \P {\mathbf{P}}
\def \R {\mathbb{R}}

\def \E {\mathbf{E}}

\def \< {\langle}
\def \> {\rangle}

\def \vx {\bm{x}}
\def \vu {\bm{u}}

\theoremstyle{plain}
\newtheorem{theorem}{Theorem}

\newtheorem{proposition}{Proposition}

\newtheorem{definition}{Definition}
\newtheorem{problem}{Problem}
\newtheorem{lemma}{Lemma}

\theoremstyle{remark}
\newtheorem{remark}{Remark}

\begin{document}
\setboolean{draft}{false}
%
\title{Distributed remote estimation over the collision channel with and without local communication}

%

\author{Xu~Zhang, Marcos M. Vasconcelos, Wei~Cui, and Urbashi Mitra
	\thanks{This work was supported in part by the following agencies: ONR under grant N00014-15-1-2550, NSF under grants CNS-1213128, CCF-1718560, CCF-1410009, CPS-1446901 and AFOSR under grant FA9550-12-1-0215.}
	\thanks{X.~Zhang and W.~Cui are with the School of Information and Electronics, Beijing Institute of Technology, Beijing 100081, China (e-mails: \{connorzx,cuiwei\}@bit.edu.cn) .}
	\thanks{M. M. Vasconcelos and U. Mitra are with the Department of Electrical
		Engineering, University of Southern California, Los Angeles, CA 90089
		USA  (e-mails: \{mvasconc,ubli\}@usc.edu).}
}


\maketitle

\begin{abstract}
The emergence of the Internet-of-Things and cyber-physical systems necessitates the coordination of access to limited communication resources in an autonomous and distributed fashion.  Herein, the optimal design of a wireless sensing system with $n$ sensors communicating with a fusion center via a collision channel of limited capacity $k\,\,(k<n)$ is considered.  In particular, it is shown that the problem of minimizing the mean-squared error subject to a threshold-based strategy at the transmitters is quasi-convex.  As such,  low complexity, numerical optimization methods can be applied. When coordination among sensors is not possible, the performance of the optimal threshold strategy is close to that of a centralized lower bound.   The loss due to decentralization is thoroughly characterized. Local communication among sensors (using a sparsely connected graph), enables the on-line learning of unknown parameters of the statistical model. These learned parameters are employed to compute the desired thresholds locally and autonomously. Consensus-based strategies are investigated and analyzed for parameter estimation. One strategy approaches the performance of the decentralized approach with fast convergence and a second strategy approaches the performance of the centralized approach, albeit with slower convergence.  A hybrid scheme that combines the best of both approaches is proposed offering a fast convergence and excellent convergent performance.
\end{abstract}

\section{Introduction}

Large-scale distributed sensor networks often face the challenge of limited bandwidth, which may lead to packet collisions in the channel in the absence of a coordination protocol among sensors\cite{kim2012cyber}. Meanwhile, sensors that are battery-powered and spread out over large areas, may lead to inefficiency in the transmission of large amounts of data to the fusion center \cite{doe2004industrial,imer2010optimal}.  In these applications, we are interested in designing systems where a large number of sensors communicate under strict bounds on allowable delay and communication bandwidth.  	Two of the main goals of 5G wireless networks is to provide reliable connectivity to a massive number of devices simultaneously and to provide communication rates able to support Artificial Intelligence applications.

Thus, both the Internet-of-Things and modern cyber-physical systems require strategies to enable the autonomous and distributed optimal allocation of limited resources. Furthermore, it is desirable that the sensors be able to operate in the absence of complete information of the underlying statistical model governing the generation of data. If possible, users should be able to operate in the  absence of a centralized authority by exchanging information locally with their neighbors.  In the presence of bandwidth constraints, a strategy is to decrease the risk of collisions and improve energy efficiency by transmitting a limited number of most informative measurements.

	\begin{figure}[t!]
		\centering
		\includegraphics[scale=0.35]{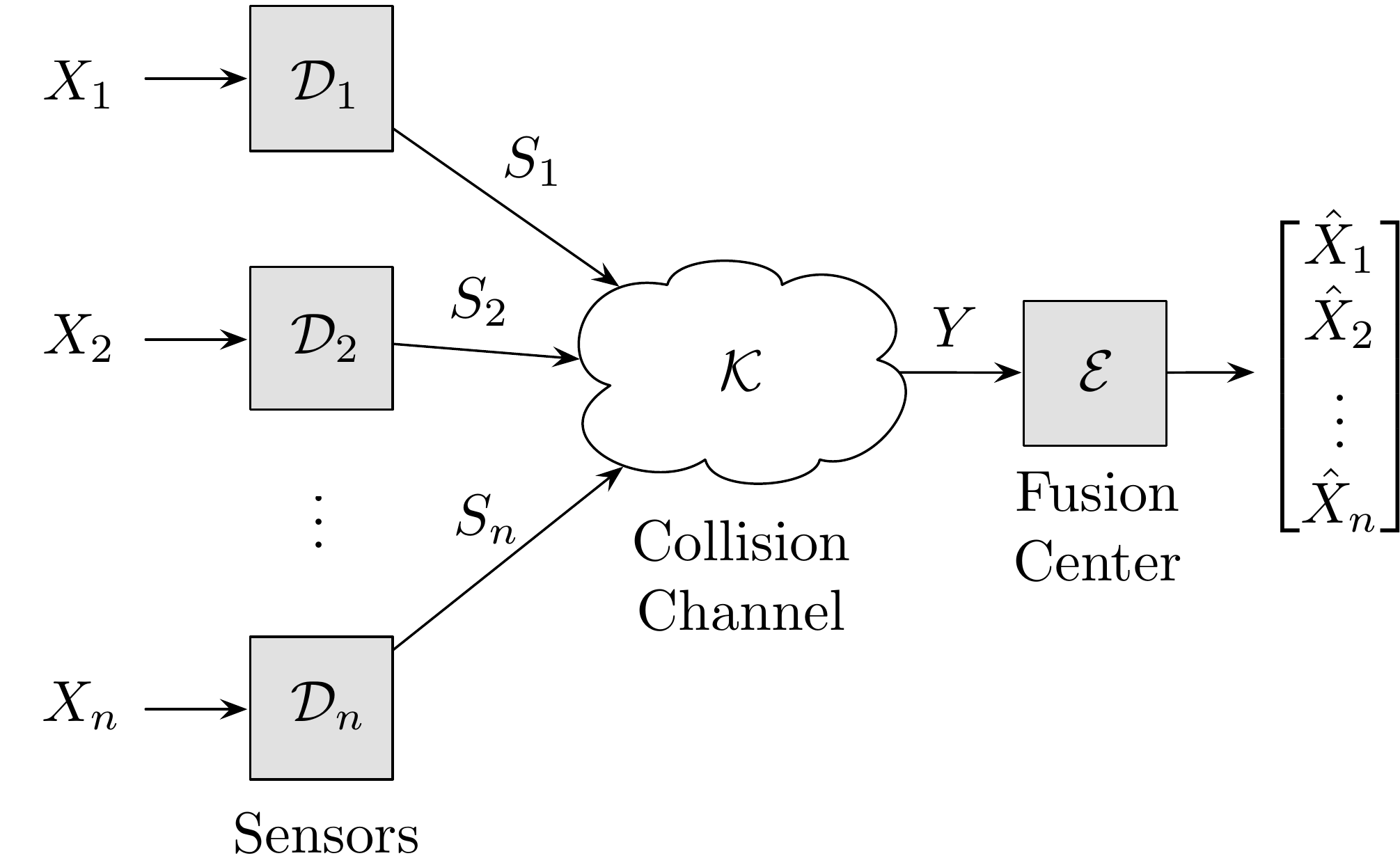}
		\caption{ System diagram for  remote estimation over the collision channel.}
		\label{fig: Systemdiagram}
	\end{figure}
	
In this paper, we study the remote sensing system depicted in \cref{fig: Systemdiagram}, where $n$ sensors observing independent and identically distributed continuous random variables communicate with a fusion center over a collision channel, which is an abstraction used to capture the effect of interference in wireless networks. The channel can only support the reliable transmission of at most $k$ packets, where $k \ll n$. If the number of simultaneous transmissions is larger than $k$, a collision occurs and is observed at the fusion center. We are interested in the design of transmission strategies to be employed by the sensors in this system that optimizes the channel access in a distributed way. We observe that a practical realization of such a system with these constraints is wireless body area sensing networks employing current implementations of wireless standards \cite{zois2014active,zois2017active}.

 Our goal is to develop new techniques for Medium Access Control (MAC) for IoT. Our abstraction for a sensor network of multiple identical sensors communicating with a fusion center over a finite capacity collision channel provides new insights for the design of alternative MAC schemes for 5G networks. For example, in many sensing applications, the communication of measurements that are uninformative can be sacrificed without significant loss in performance, freeing resources to the remaining sensors in the network. This cooperation among sensors is the centerpiece of this article, which seeks to lay the foundations of a new framework for distributed MAC protocols under assumptions of the probabilistic model of the observations.

	\subsection{Related Literature}
	
    The optimal design of remote estimation systems has been of great interest in the past two decades. There exists a rich literature on these systems under different technical assumptions.
	
	Remote sensing with a single sensor was considered in \cite{imer2010optimal,lipsa:2011,Nayyar:2013,Leong:2018,Leong2018Transmission,Leong2019Information,Gao2019Communication,Lu2020Optimal,Ding2020Defensive}. A dynamic system with a single sensor under a limited number of transmissions was studied in \cite{imer2010optimal} and the optimal strategy under the MSE criterion was obtained. Instead of limiting the number of transmissions, the authors in \cite{lipsa:2011} considered  costly communication, which showed that a symmetric threshold transmission strategy and a Kalman-like estimator jointly optimize the problem. 
	Assuming that the single sensor harvests energy randomly from its environment and uses this energy for communication, optimal strategies to minimize the estimation error  were provided in \cite{Nayyar:2013,Leong:2018}. Optimal strategies for remote estimation in the presence of attacks were designed in \cite{Leong2018Transmission,Leong2019Information,Gao2019Communication,Lu2020Optimal,Ding2020Defensive}. 
	
	
	Remote sensing with multiple sensors sharing the same communication channel was studied to allocate limited resources optimally. In presence of the feedback from fusion center, transmission strategies were proposed in \cite{huang2006dynamic,dogandzic2008decentralized,michelusi2015cross1,michelusi2015cross2, Wu2018Optimal,Li2018Transmit,Ni2019Pricing,Redder2019Deep,Wu2019Learning,Leong2020Deep}. 
	The feedback is used in two ways, one acting as the control signals while the other providing summary information to sensors. In \cite{Wu2018Optimal}, a multi-sensor system over shared channel with limited packet sizes  was studied, where sensors make local  decisions informed by a control signal from the fusion center. In \cite{huang2006dynamic,dogandzic2008decentralized}, a Bayesian framework is developed for decentralized remote sensing, which adapts the quantization rate  of transmissions based on  feedback from fusion center. In \cite{michelusi2015cross1,michelusi2015cross2}, the fusion center feeds back estimation quality that is coupled with local sensor quality to jointly optimize cross-layer performance. Combined with reinforcement learning, several works designed  optimal strategies for more realistic systems with feedback \cite{Redder2019Deep,Wu2019Learning,Leong2020Deep}. 
	
	
In contrast remote estimation without feedback was examined in \cite{Weerakkody2016Multi,Han2017Optimal,DING20173829,pezzutto2020transmission,vasconcelos:2017a,vasconcelos2018optimal, Zhang:2020}. In \cite{DING20173829},   multi-sensor remote estimation over a shared channel by using correlated equilibrium is studied, where transmissions from other sensors are viewed as interference. In \cite{pezzutto2020transmission}, collision resolution for two sensors is considered with improvements over orthogonal transmission schemes.
	Previous work \cite{vasconcelos:2017a} examined a multi-sensor system under one-shot transmission over collision channel and showed that there are asymmetric threshold strategies that are optimal under the MSE criterion. A similar system with discrete random variables under one-shot transmission is considered in a more recent paper \cite{vasconcelos2018optimal}.  In contrast to this prior work, herein we study a multi-sensor system that supports multiple ($k>1$) simultaneous packets. To achieve improved performance, we also consider the case where local communication among sensors is allowed.

	
Finally, the problem considered in the paper has a close relationship with the problem of observation-driven scheduling for remote sensing  in a one-shot transmission studied in \cite{vasconcelos2019observation}, in which a scheduler collects the measurements from all sensors, and chooses a single one to be transmitted to the destination. By jointly designing the scheduling and estimation policies, the scheme in  \cite{vasconcelos2019observation} sends the largest measurement to the fusion center and avoids the collision in the channel, which can be regarded as a centralized version of the approach proposed herein.

Based on \cite{vasconcelos2019observation},  a bandwidth constraint of $k$ necessitates the need to determine the sensors with the $k$ largest measurements. Learning the top $k$ measurements in a decentralized fashion has been considered in \cite{madden2002tag,wu2007top,yeo2009data, malhotra2010exact,liao2012efficient}. A logical tree topology was used to aggregate the measurements in \cite{madden2002tag}. A filter-based approach was studied in \cite{wu2007top}. A grid-based method was considered in \cite{liao2012efficient}. Thus we see that data aggregation and feedback are typically employed to learn the top $k$ measurements.  In contrast, we develop a threshold-based decentralized method via consensus, where each sensor estimates the $k$-th largest measurement as the threshold and determines whether to send.  A purely disributed scheme is achieved via consensus to learn quantiles of observed data.

	\subsection{Contributions}
	
    The main contributions of this paper are:
    
\begin{itemize}
	
	\item In the absence of local communication among sensors, we study the design of a globally optimal threshold communication strategy under a symmetry assumption of the probability distribution of the observations. Our analysis shows that under this assumption, the mean-squared error is a strictly quasi-convex function of the threshold, which is amenable to low complexity numerical optimization schemes. 
    This result is valid for any probability function that is symmetric around the mean of the observed random variable. More importantly, our results guarantee the existence of a single optimal threshold.
			
	\item In the presence of local communication, the sensors can coordinate to learn a common threshold strategy when the underlying probability distribution is unknown or not completely specified. In this case, there is a trade-off between performance and delay before the decision at the sensors is taken.
	
	    \begin{enumerate}			
			
			\item When the information at the sensors about the distribution is incomplete, we propose an approach based on consensus, where each sensor estimates the unknown parameter(s) based on its local observation and information received from its neighbors. Subsequently, each sensor computes its threshold and uses it to determine locally whether to attempt transmission or not.   
			
			\item When the distribution is unknown, we propose a distributed quantile regression method, where each sensor estimates the $k$-th largest observation among all sensors, and uses it as the threshold to decide whether or not to transmit. We show that this scheme approaches the performance of the optimal centralized scheme. However, it has a slow convergence rate, due to having to learn more parameters than in the threshold based scheme above. 
			
			\item Finally, if the distribution is partially known, we propose a scheme that initially uses the algorithm based on consensus to bootstrap the scheme based on quantile regression. This scheme achieves both fast convergence and asymptotic performance close to the centralized optimal. We provide an example which shows that the scheme is robust to mismatch in the underlying assumptions, i.e. one can assume Gaussian distribution when the true distribution is not Gaussian.  

		\end{enumerate}

\end{itemize}

	\section{Problem Formulation} \label{sec: Problem Formulation}	
	
	In this section, we establish the problem setup for decentralized remote estimation system over a collision channel of capacity $k$. Consider the system diagram shown in \cref{fig: Systemdiagram}. There are $n$ sensors and a fusion center $\mathcal{E}$, which are connected by a collision channel $\mathcal{K}$. The $i$-th sensor observes a zero-mean random variable $X_i$, $i\in\{1,\cdots,n\}$. The random variables $\{X_i\}_{i=1}^{n}$ are independent and identically distributed (i.i.d.), and admit a probability density function (pdf) $f_X(x)$,
such that $f_X(x)>0$ for $x \in \R$.
	Each sensor decides whether to transmit its observed measurement to the fusion center or to remain silent according to a threshold strategy defined as follows. 
	
	\begin{definition} [Threshold strategy]
		Let $D_i \in \{0,1\}$ be the binary decision variable of the $i$-th sensor, where $D_i=1$ denotes that the sensor decides to transmit its measurement, and $D_i=0$ denotes that the sensor decides to remain silent.	
		A threshold strategy for the $i$-th sensor is a function $\mathcal{D}_i: \mathbb{R}\rightarrow \{0,1\}$ such that 
		\begin{equation}
		\mathcal{D}_i(x) \triangleq \mathbf{1}(|x| \ge T), \label{eq:threshold_strategy}
		\end{equation}
		where $T \in [0,+\infty)$ denotes the threshold and  $\mathbf{1}(\mathfrak{S})$ denotes the indicator function of the statement $\mathfrak{S}$.
	\end{definition}
	
	\begin{remark} This formulation is an instance of a symmetric stochastic team. This special type of team decision problems is often more tractable because the optimization problem is over a single parameter ($T$) and it also allows for studying the dependence of the results with respect to the number of sensors. In some cases, it is possible to study the performance of the system in the regime when the number of sensors is infinite, which is particularly relevant to IoT applications.
	\end{remark}

	
	After making a decision, each sensor produces a channel input packet, $S_i$, defined as follows:
	\begin{equation}
	S_{i} \triangleq \left\{\begin{array}{ll}{(i,X_{i})} & {\text { if } D_i=1} \\ {\varnothing} & {\text { if } D_i=0}\end{array}\right., \quad i \in\{1,\ldots,n\}.
	\end{equation}
	\begin{remark}
		We assume that if a sensor decides to transmit, its unique identification number $i$ is transmitted along with its measurement. This is done so that the receiver can identify the origin of the successfully received communication packets without ambiguity.
	\end{remark}
	The collection of $n$ sensors share a collision channel $\mathcal{K}$ of limited capacity $k$, defined as follows:
	
	\begin{definition}[Collision channel of capacity $k$] 
		The collision channel of capacity $k$ allows the communication of at most $k\leq n$ simultaneous packets. Let $\mathbb{D}\triangleq\{i \mid D_i=1\}$ denote the set of indices of all transmitting sensors.
		The output of the collision channel $Y$ is given by: 
		\begin{equation}
		Y \triangleq
		\begin{cases}
		\varnothing & \text{if} \ \ |\mathbb{D}| = 0 \\
		\big\{(i, X_{i}) \mid i\in \mathbb{D}\big\} & \text {if}  \ \ 1 \leq |\mathbb{D}|\leq k\\  
		\(\mathfrak{C},\mathbb{D}\) &  \text{otherwise}.
		\end{cases}
		\end{equation}
		The special symbol ${\mathfrak{C}}$ denotes that a collision occurred and $\varnothing$ denotes that the channel is idle.
	\end{definition}
	\begin{remark} \label{rm:indicies collision}
		When a collision occurs, we assume that the fusion center can decode the indices of the transmitting sensors. This assumption is realistic, specially for 5G networks which have very large bandwidth and allow for advanced coding and signal processing techniques \cite{1469098,7967038}. 
	\end{remark}

	
	Our purpose is to solve the following estimation problem over the collision channel under the normalized mean squared error (MSE) criterion.
	\begin{problem}  \label{prl: main} 
		Assuming that each sensor uses a threshold strategy of the form given in Eq.~(\ref{eq:threshold_strategy}), given the number of sensors, $n$, the pdf of the sensors' observations, $f_X$, and the capacity of the collision channel, $k$; find a threshold $T$ that minimizes
		\begin{equation} \label{eq: MSE}
		\mathcal{J}_{n,k}(T) \triangleq \frac{1}{n}\E\[\sum_{i=1}^{n}\(X_i-\hat{X}_i\)^2\],
		\end{equation}
		where the estimates $\hat{X}_i$ are given by:
		\begin{equation}\label{eq:optimal_estimates}
		\hat{X}_i = \mathbf{E}[X_i \mid Y], \ \ i \in\{1,\cdots,n\}.
		\end{equation}
	\end{problem}
	
	\begin{remark} Due to the definition of the collision channel, any estimate $\hat{X}_i$ depends on the entire set of decision variables $\{D_i\}_{i=1}^{n}$. Such estimate will be derived in the following section. 
	\end{remark}


	\section{Optimal decentralized scheme without local communication} \label{sec: no local communication}
	
	\subsection{Quasi-convexity of \cref{prl: main}}
	Assuming that there is no local communication among the sensors, and that the distribution of the observations is symmetric, we will provide a solution for Problem \ref{prl: main}. In particular, we begin by providing alternative expressions for Eqs. \eqref{eq: MSE} and \eqref{eq:optimal_estimates}. We will then show the quasi-convexity of \cref{prl: main}, which can thus be solved using simple numerical procedures. 
	
	\begin{lemma} \label{lm: estimator2}
		Provided the pdf $f_X$ is symmetric, given the decision variables $\mathbb{D}= \{i \mid D_i=1\}$, the output of the estimator can be rewritten as
		\begin{equation} \label{eq: estimator1}
		\hat{X}_i \stackrel{\mathrm{w.p.1}}{=} \begin{cases}
		X_i &  \text{if} \ \ |\mathbb{D}|\leq k \ \text{and} \ i\in\mathbb{D},\\
		0 & \text{otherwise}. 
		\end{cases}
		\end{equation}
		for $i \in \{1,\cdots,n\}$.
	\end{lemma}
	\begin{IEEEproof}
		We compute the conditional expectation in \cref{eq:optimal_estimates} for every possible output of the collision channel.
		
		When there is no collision and $X_i$ was transmitted, i.e., $ |\mathbb{D}|\le k$ and  $D_i=1$,  we have $(i,X_i) \in Y$, which implies that
		\begin{equation}
		\hat{X}_i=\E[X_i \mid Y]\stackrel{\mathrm{w.p.1}}{=}X_i.
		\end{equation}
		
		When a collision occurs and  $X_i$ was transmitted, i.e., $ |\mathbb{D}| > k$ and  $D_i=1$,  we have $Y=(\mathfrak{C},\mathbb{D})$ and know $i \in \mathbb{D}$ from Remark \ref{rm:indicies collision}, which implies that
		\begin{equation}
		\hat{X}_i  \stackrel{(a)}{=}  \E[X_i \mid D_i=1]  =  \E[X_i \mid |X_i|\geq T]  \stackrel{(b)}{=}  0, 
		\end{equation}
		where $(a)$ is due to $\{X\}_{i=1}^{n}$ being a collection of independent random variables, and $(b)$ is due to the symmetry of the pdf $f_X$.
		
		When $X_i$ is not transmitted, i.e., the index $i$ does not appear in the channel output $Y$, which implies $D_i=0$. In this case, we have 
		\begin{equation}
		\hat{X}_i   =  \E[X_i \mid D_i = 0 ] = \E[X_i \mid |X_i| < T]  \stackrel{(c)}{=} 0,
		\end{equation}
		where $(c)$ is due to the symmetry of the pdf $f_X$.
		
	\end{IEEEproof}
	
	\begin{lemma} \label{lm: cost function2}  Let $\{X_i\}_{i=1}^n$ be an i.i.d. sequence distributed according to a symmetric pdf $f_X$. The objective function in Problem \ref{prl: main} can be expressed as:
		\begin{equation} \label{eq: lossfunction} 
		\mathcal{J}_{n,k}(T)=\E\[X^2\]- \E\[X^2 \mathbf{1}(|X| \ge  T)\] F_{n,k}(T),
		\end{equation}
		where
		\begin{equation} \label{OrderStatistics_cdf}
		F_{n,k}(T) \triangleq \sum_{\ell=0}^{k-1} \binom{n-1}{\ell}\big(1-p(T)\big)^{\ell} p(T)^{n-1-\ell},
		\end{equation}
		and 
		\begin{equation}
		p(T) \triangleq \mathbf{P}(|X|< T).
		\end{equation} 
	\end{lemma}
	\begin{IEEEproof}
		See Appendix \ref{appendix: cost function2}.
	\end{IEEEproof}

	\begin{theorem} \label{thm: optimality}
The cost function $\mathcal{J}_{n,k}(T)$ in \cref{eq: lossfunction} is strictly quasi-convex and admits a unique optimal threshold $T^{\star}$ such that
	\begin{equation} \label{eq: optimality}
		T^\star=\arg \min_{T \geq 0} \mathcal{J}_{n,k}(T).
		\end{equation}
	\end{theorem}
	\begin{IEEEproof}
		See Appendix \ref{appendix: optimality}.
	\end{IEEEproof}



\begin{remark} The result in \cref{thm: optimality} holds for any symmetric pdf, regardless of the number of modes of the distribution. We highlight that proving quasi-convexity is typically a non-trivial task and existing methods rely on composition rules of operations that preserve quasi-convexity, which are not available in our case. From an algorithmic standpoint, quasi-convexity is a property as desirable as convexity. Although a closed-form expression to $T^\star$ is unlikely to exist, we can compute it via iterative numerical methods. Due to the continuity and quasi-convexity of $\mathcal{J}_{n,k}(T)$ (established in Appendix \ref{appendix: optimality}), we can use numerical methods from disciplined quasi-convex programming to compute the optimal threshold \cite{agrawal2020disciplined}.
\end{remark}


	

When using numerical optimization solvers, it is important to properly initialize the interval to be searched, especially when the support of the pdf $f_X$ is unbounded.
Next, we will provide an interval initialization by analyzing the $0-1$ phase transition property of
\begin{equation}
	F_{n,k}(T)=\P \big( \text{at most $k-1$ out of $n-1$ sensors} \  \text{decide to transmit}\big).
\end{equation} 
	By inspection of \cref{eq: lossfunction}, when $T$ is such that $F_{n,k}(T) \approx 0$, the cost is $\mathcal{J}_{n,k}(T) \approx \E[X^2]$; when $T$ is such that $F_{n,k}(T) \approx 1$, then 
\begin{equation}
\mathcal{J}_{n,k} \approx \mathbf{E}\big[X^2\mathbf{1}\big(|X|<T\big)\big],
\end{equation}
which is non-decreasing in $T$. Therefore, the optimal $T^\star$ should occur in the interval when $F_{n,k}$ transitions from $0$ to $1$.

	\begin{lemma} \label{lm: initialization}
		Let $T^\star$ be the optimal threshold for the cost function $\mathcal{J}_{n,k}(T)$ in \cref{eq: lossfunction}. Then 	
		\begin{equation}
		    T^\star \geq p^{-1}\(1-\frac{k}{n}\).
		\end{equation}
	\end{lemma}

	\begin{IEEEproof}
	See Appendix \ref{Appendix: Proof_Lemma_initialization}.
	\end{IEEEproof}

		
	\begin{lemma} \label{lm: phasetransition} 
		Let $s>0$. Then
			\begin{equation}
			    F_{n,k}(T)  \ge  1- e^{-s^2}, \ \ T > p^{-1}\(1 - \frac{k-s\sqrt{2k}}{n-1}\).
			\end{equation}
	\end{lemma}
	
	\begin{IEEEproof}
		See Appendix \ref{Appendix: phasetransition}.
	\end{IEEEproof}

\begin{theorem}\label{thm:initial_interval}
There exists $\bar{s} > 0$ such that:
\begin{equation}
    p^{-1}\(1-\frac{k}{n}\) \leq T{^\star} \leq p^{-1}\(1 - \frac{k-\bar{s}\sqrt{2k}}{n-1}\).
\end{equation}
\end{theorem}

\begin{IEEEproof}
The proof follows from \cref{lm: initialization,lm: phasetransition}.
\end{IEEEproof}

\begin{remark}
\Cref{thm:initial_interval} provides an interval that is guaranteed to contain the optimal solution. Moreover, by using the result in \Cref{thm:initial_interval}, we can avoid initializing the numerical solver where $\mathcal{J}_{n,k}(T)$ is flat, which may lead to falsely declare that a local minimum has been found, and failing to find the unique global minimum guaranteed by \cref{thm: optimality}.
\end{remark}


%
%
%
	
	
	\subsection{A centralized lower bound to \cref{prl: main}} \label{sec:lowerbound}
	
	When the goal is to minimize the MSE of zero-mean independent variables such as in Problem 1, the optimal centralized strategy consists of transmitting the $k$ largest measurements in magnitude to the fusion center \cite{vasconcelos2019observation}. The performance of this strategy serves as a lower bound to decentralized communication strategies over the collision channel with capacity $k$. For the ``top-$k$'' strategy, the value of the cost function is given by 
	\begin{IEEEeqnarray}{rCl} 
	\mathcal{J}_{n,k}^L & \triangleq & \frac{1}{n}\sum_{i=k+1}^{n}\E\big[ Z_{(i)}^2\big] 
	\label{eq:lowerbound}
	\end{IEEEeqnarray}
	where  $Z_{i} \triangleq |X_i|$, and $Z_{(i)}$ is defined as the $i$-th largest value in $\{Z_\ell\}_{\ell=1}^{n}$ such that:
	\begin{equation}
Z_{(n)} \leq Z_{(n-1)} \leq \cdots \leq Z_{(1)}.
	\end{equation}

	From results on ordered statistics \cite{arnold2012relations}, the second moment of $Z_{(i)}$ is given by:
	\begin{equation}\label{eq: moment of Z_(i)}
			\E\big[Z_{(i)}^2\big] = \frac{\int_{0}^{\infty}z^2F_Z(z)^{n-i}\big(1-F_Z(z)\big)^{i-1}f_Z(z)dz}{\mathrm{B}(n-i+1,i)},
	\end{equation}
	where $f_Z$ and $F_Z$ are the pdf and cdf of $Z$, respectively, and $\mathrm{B}(\cdot,\cdot)$ denotes the beta function. Since $Z=|X|$, we have:
	\begin{equation}
	F_Z(z) = 2F_X(z) -1, \ \ z\geq 0,
	\end{equation}
	and
	\begin{equation}
	f_Z(z) = 2f_X(z), \ \ z\geq 0.
	\end{equation}

	This lower bound is used as benchmark in the examples shown in this paper. The gap between the performance of the optimal threshold policy and the value of $\mathcal{J}^L_{n,k}$ corresponds to the loss due to decentralization.
	

	\subsection{Numerical results}
	
	\begin{figure}[!t]
		\centering
			\includegraphics[width=0.6\linewidth]{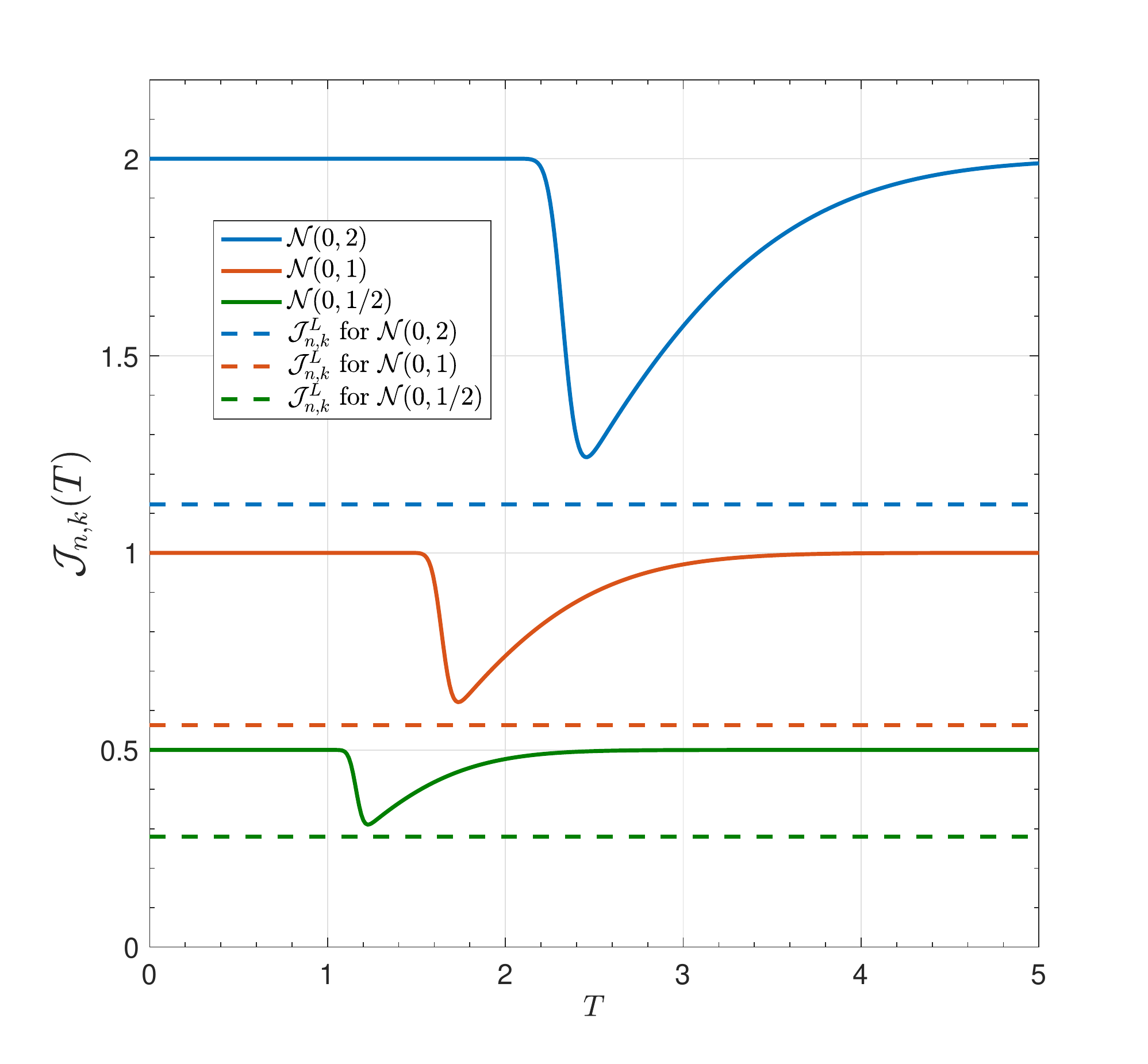}
		\caption{Cost function $\mathcal{J}_{n,k}(T)$ as a function of the threshold $T$ with $n=1000$ sensors and a collision channel with capacity $k=100$ packets for Gaussian observations of different variances. The dashed horizontal lines represents the corresponding centralized lower bounds $\mathcal{J}_{n,k}^L$ given in \cref{eq:lowerbound}.}
		\label{fig: CostJvsT}
	\end{figure}
	
	
	\Cref{fig: CostJvsT} shows the normalized MSE $\mathcal{J}_{n,k}(T)$ for a system with $n=1000$ sensors and a collision channel of capacity $k$ making Gaussian observations with different variances. We can observe the quasi-convexity property, and compare the performance of the optimal decentralized scheme $\mathcal{J}_{n,k}(T)$ to the centralized lower bound $\mathcal{J}^L_{n,k}$. From this figure, we can also observe that $\mathcal{J}_{n,k}(T)$ is flat at regions away from the optimal threshold $T^\star$. This observation reinforces the need for \cref{thm:initial_interval} and proper initialization of the numerical solvers used to compute $T^\star$. 
	
	 
			\begin{figure}[!t]
	    \centering
		\includegraphics[width=0.6\linewidth]{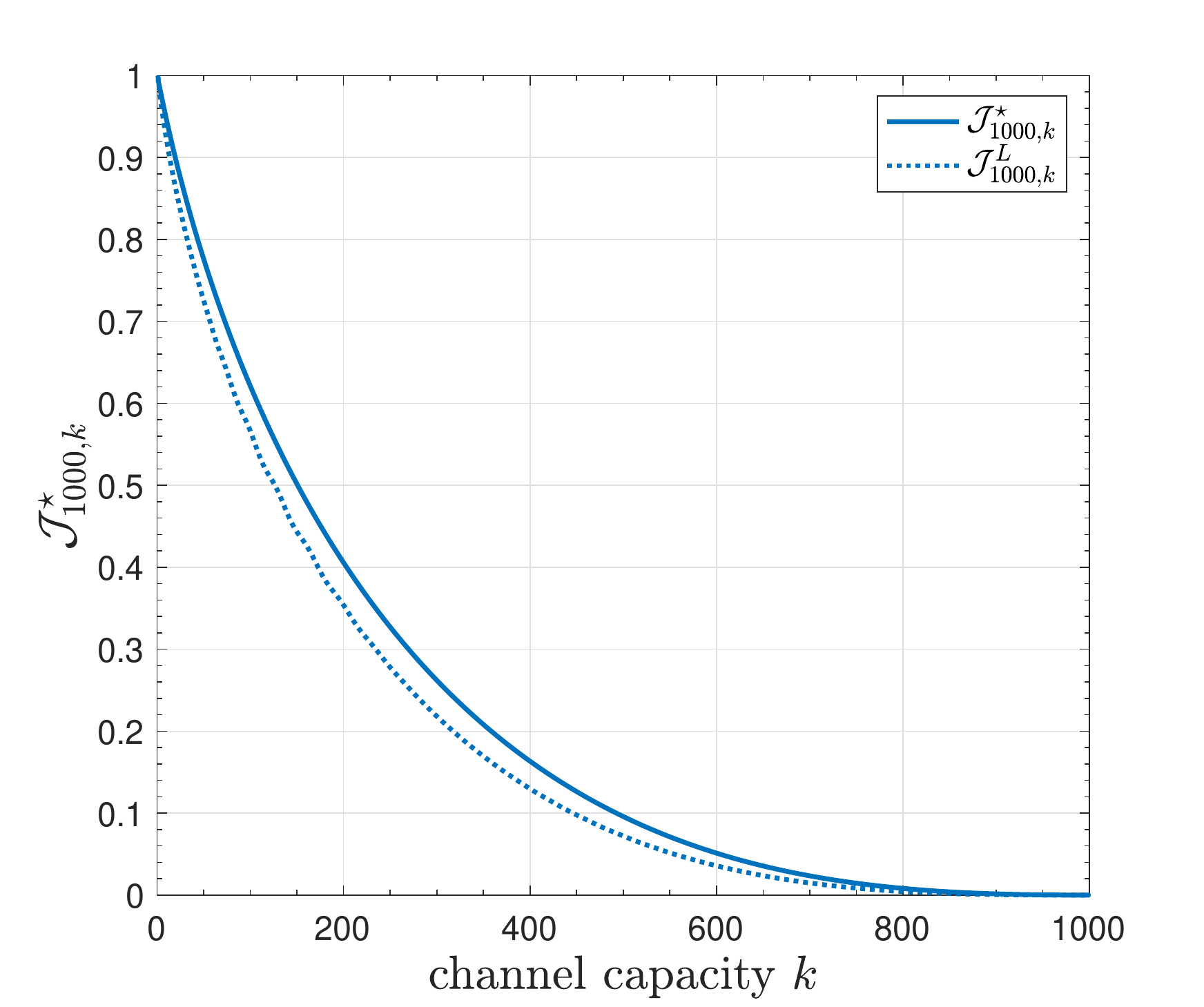}
		\caption{Optimal cost $\mathcal{J}_{n,k}(T^\star)$ and the lower bound $\mathcal{J}_{n,k}^L$ as functions of  the capacity of the collision channel $k$ (performance of the decentralized and centralized schemes) with $n=100$. The observations at the sensors are i.i.d. according to a standard Gaussian distribution, $X\sim \mathcal{N}(0,1)$.}
		\label{fig: OptimalJvsK}
	\end{figure}

	

	 For a system with $n=1000$ sensors, \cref{fig: OptimalJvsK} displays the dependency of the optimal MSE $\mathcal{J}_{n,k}(T^\star)$ and the lower bound $\mathcal{J}_{n,k}^{L}$ as function of the capacity of the collision channel $k$ for standard Gaussian observations, $X_i\sim \mathcal{N}(0,1)$. As the capacity $k$ increases, more measurements are successfully received at the fusion center, and the normalized MSE decreases. 
	We can also observe that the optimal choice for the threshold successfully mitigates the occurrence of collisions. Consequently, the decentralized scheme performs reasonably close to the centralized scheme. The difference between the solid (decentralized) and the dotted (centralized) curves is the performance loss due to decentralization.

\section{Decentralized schemes with local communication} \label{sec: local communication}

	
	
	Consider a connected undirected graph $\mathbb{G}=\(\mathbb{N},\mathbb{E}\)$ with $n$ nodes, each node represents a sensor observing an independent random variable as before. Here, $\mathbb{N}=\{1, \ldots, n\}$ denotes the set of sensors and $\mathbb{E} \subset \mathbb{N} \times \mathbb{N}$ denotes the set of edges between nodes. 
	Let $\mathbb{N}_i$ denote the set of neighbors of the $i$-th sensor, and $d_{\max}\triangleq \max_{i} |\mathbb{N}_i|$. By local communication, we mean that if $(i,j)\in\mathbb{E}$, sensors $i$ and $j$ can communicate with each other for a given number of rounds, before making their final decisions on whether attempt a transmission to the fusion center or not. Each round of communication represents one unit of accrued delay in communication between the sensors and fusion center.
	

	\subsection{Consensus-based decentralized scheme}\label{sec: consensus}

	In many scenarios, we may not have access to one or more parameters of the pdf $f_X$ although we know that the distribution is of a certain type, e.g. we may know that the distribution is Gaussian, but its variance is unknown. By means of local communication among the sensors, we enable them to estimate the unavailable parameters in a distributed way, such that the optimal threshold $T^\star$ may be computed in a decentralized way. This is done at the expense of some delay in communication with the fusion center. We use a consensus scheme \cite{xiao2004fast,nedic2010constrained} to estimate the unknown parameters of the distribution. We will illustrate how the method works for the Gaussian distribution where $X \sim \mathcal{N}(0,\sigma^2)$.
	
	
	Let $y_i(t)$ denote the local estimate of the variance of the $i$-th sensor at the $t$-th round of local communication. We initialize the local estimates by setting $y_i(0)=x_i^2$, $i=\{1,\cdots,n\}$. On the $t$-th round of local communication each sensor performs the following steps:
	
	\begin{enumerate}
		\item \textbf{Distributed variance estimation:} Each node updates its local estimate based on the local estimates of its neighbors according to:
		\begin{equation} \label{eq: AverageConsensus}
		y_i(t+1) = y_i(t) + \frac{1}{d_{\max}}\sum_{j\in \mathbb{N}_{i}}\big(y_j(t)-y_i(t)\big),
		\end{equation}
		for $i \in\{1,\cdots,n\}$.
		\item \textbf{Threshold computation}. Using the techniques introduced in Section 3 and assuming that $X\sim \mathcal{N}(0, y_i(t))$, each node solves: 
\begin{equation} \label{eq:local_threshold}
T_i^{\star}(t) = \arg\min_{T\geq 0} \mathcal{J}_{n,k}(T),
\end{equation}
where $\mathcal{J}_{n,k}(T)$ is given by \cref{eq: lossfunction}. 

	\end{enumerate}

If at time $t$ the sensors use the thresholds $\{T_i^\star(t)\}_{i=1}^n$, the decision variables $u_i(t)$ are computed as:
\begin{equation}
u_i(t) = \mathbf{1}\big(|x_i| \geq T_i^\star(t)\big),
\end{equation}
and the instantaneous performance of this approximate scheme is given by
\begin{equation}
\mathcal{J}_C\big(\boldsymbol{x},\boldsymbol{u}(t)\big)\triangleq
		\begin{cases}
		\frac{1}{n}\sum \limits_{i=1}^{n} x_i^2\big(1- u_i(t)\big), &  \text{if} \ \ \sum \limits_{i=1}^{n}u_i(t) \leq k,\\ 
		\frac{1}{n}\sum \limits_{i=1}^{n}x_i^2, & \text {if} \ \ \sum \limits_{i=1}^{n} u_i(t) > k.
		\end{cases}
		\end{equation}



\begin{figure}[!t]
	\centering
		\includegraphics[width=0.6\linewidth]{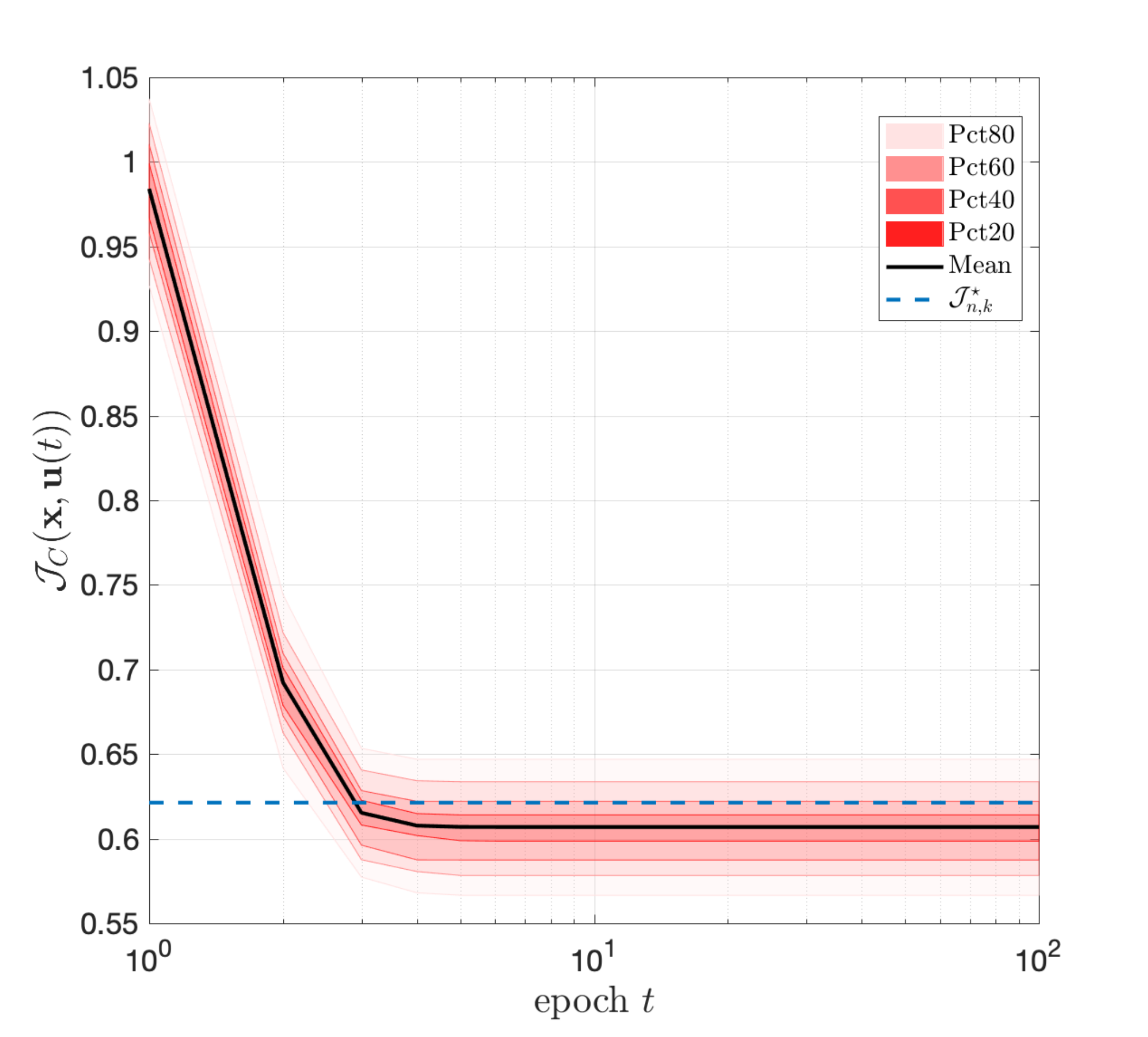}
		\caption{Plot representing the mean and percentiles of 100 sample paths $\mathcal{J}_C\big(\boldsymbol{x},\boldsymbol{u}(t)\big)$ for a system with $n=1000$ sensors and channel capacity $k=100$.}
		\label{fig: performance_consensus.pdf}
	\end{figure}

In \cref{fig: performance_consensus.pdf}, we simulate the performance of this scheme by generating $100$ independent sample paths and plotting the mean and different percentiles. The underlying graph $\mathbb{G}$ is a sample from the Erdos-Renyi ensemble of random graphs with edge probability $p_{\text{edge}}=0.05$, $d_{\max}=81$ and $\lambda_2=27.35$, where $\lambda_2$ is the second largest eigenvalue of the graph Laplacian, a standard measure of connectivity. One key observation here is that the mean of the sample paths converge to a value below the performance of the optimal scheme $\mathcal{J}^\star_{n,k}.$ The reason why this is the case is that the empirical average of asymptotic performance of the sample paths is always a downward biased estimator of the true optimal solution of the stochastic optimization problem $\mathcal{J}^{\star}_{n,k}$ \cite{shapiro2014lectures}.

	\subsection{Quantile-based decentralized scheme}\label{sec:quantile-based}


	When local communication among sensors is available, nothing prevents the sensors to coordinate and attempt to implement the optimal centralized scheme. Therefore, one possibility consists in each sensor keeping a local estimate of $z_{(k)}$ and using this estimate as a threshold. If the estimates are perfect, only the sensors holding the measurements with the $k$ largest magnitudes will transmit. Let $w_i(t)$  denote the estimate of $z_{(k)}$ for the $i$-th sensor at $t$-th iteration.




	We will use a distributed subgradient method to estimate the sample quantile based on \cite{7914788} corresponding to the desired ordered statistics. Let 
\begin{equation}
	\boldsymbol{z} = \big[|x_1|,|x_2|,\cdots,|x_n|\big]
	\end{equation}
	and define the following empirical cdf
	\begin{equation}
\widehat{F}(\xi ; \boldsymbol{z})\triangleq \frac{1}{n} \sum_{i=1}^{n} \mathbf{1}( z_i \leq \xi).
	\end{equation}

	Let the sample quantile be defined as
	\begin{equation}
	\theta_{p}\triangleq \inf \Big\{\xi \mid \widehat{F}(\xi ; \boldsymbol{z}) \geq p \Big\}.
	\end{equation}
	

\begin{proposition}[Relationship between sample quantiles and ordered statistics]
Let $\{z_i\}$ be a sequence of realizations of the i.i.d. sequence of continuous random variables $\{Z_i\}$ and its corresponding reordering $\{z_{(t)}\}$. If 
\begin{equation}
p \in \Big(\frac{n-k}{n},\frac{n-k+1}{n}\Big),
\end{equation}
then 
\begin{equation}
\theta_{p}=z_{(k)}.
\end{equation}
\end{proposition}


	
Let $w_i(0)=z_i$, $i\in\{1,\cdots,n\}$ and $\psi_i(t)$ the message sent by the $i$-th sensor at the $t$-th iteration to its neighbors. Let $\eta(t)$ be a deterministic step-size sequence, which is chosen as: 
\begin{equation}\label{eq:step}
\eta(t)=\frac{\alpha}{t^{\tau}} 
\end{equation}
where $\alpha$ is a positive constant and $\tau \in \textcolor{black}{(0.5,1]}$.
	
On the $t$-th round of local communication we perform the following steps:

	\begin{enumerate}
	
	\item \textbf{Message computation:}
	\begin{equation}
		\psi_i(t)=w_i(t)-\eta(t) s_{i}\big(z_i,w_i(t-1)\big),
	\end{equation}
	where
	\begin{equation}
	 s_{i}\big(z_i,w_i(t-1)\big) \triangleq \begin{cases}
	 -\frac{p}{n} & z_i > w_i(t-1) \\
	 \frac{1-p}{n} & z_i < w_i(t-1) \\
	 0 & z_i = w_i(t-1)
	 \end{cases}
	\end{equation}
	
	\item \textbf{Local estimate update:}
	\begin{equation}
	w_i(t) = \sum_{j=1}^n c_{ij}\psi_j(t),
	\end{equation}
	where
	\begin{equation}
	c_{ij} \triangleq \begin{cases}
	\frac{1}{\max\{|\mathbb{N}_i|,|\mathbb{N}_j|\}} & j \in \mathbb{N}_i\backslash i \\
	1 - \sum \limits_{\ell \in \mathbb{N}_i\backslash i} c_{i\ell} & j=i \\
	0 & j \notin \mathbb{N}_i.
	\end{cases}
	\end{equation}

	\end{enumerate}

	If at time $t$ the sensors use the thresholds $\{w_i(t)\}_{i=1}^n$,  the decision variables $u_i(t)$ are computed as:
\begin{equation}
u_i(t) = \mathbf{1}\big(|x_i| \geq w_i(t)\big),
\end{equation}
and the instantaneous performance of this scheme is given by
\begin{equation}
\mathcal{J}_Q\big(\boldsymbol{x},\boldsymbol{u}(t)\big)\triangleq
		\begin{cases}
		\frac{1}{n}\sum \limits_{i=1}^{n}x_i^2\big(1- u_i(t)\big), &  \text{if} \ \ \sum \limits_{i=1}^{n}u_i(t) \leq k, \\ 
		\frac{1}{n}\sum \limits_{i=1}^{n} x_i^2, & \text {if} \ \ \sum \limits_{i=1}^n  u_i(t) > k.
		\end{cases}
		\end{equation}



    \begin{theorem} \label{thm: QuantileCovergence}
        Let $p \in \big((n-k)/n,(n-k+1)/n\big)$. Then,
    \begin{equation} \label{eq: QuantileCovergence}
	\lim_{t \to +\infty} w_i(t)=z_{(k)}, \ \ i \in \{1,\ldots,n\}.
    \end{equation}
    \end{theorem}
    \begin{IEEEproof}
    This result is a corollary of Theorem 1 in \cite{7914788}.
    \end{IEEEproof}
    
    One consequence of \cref{thm: QuantileCovergence} is that for a large enough delay in communication, the performance of the scheme based on sample quantile estimation converges to the bounded interval, which is specified by the following result. 
    
	\begin{theorem} \label{thm: convergence_Q} Let $p \in \big((n-k)/n,(n-k+1)/n\big)$. There exists a number $M>0$ such that for $t \geq M$,
		\begin{equation}
			\frac{1}{n}\sum_{i=k+1}^{n} z_{(i)}^2 \le \mathcal{J}_Q\big(\vx,\vu(t) \big)\le \frac{1}{n}\sum_{i=k}^{n} z_{(i)}^2.	
		\end{equation}
	\end{theorem}	
		\begin{IEEEproof}
	From Theorem \ref{thm: QuantileCovergence}, we have:
\begin{equation}
	\lim_{t \to +\infty} w_i(t)=z_{(k)}, \ \ i \in \{1,\ldots,n\}.
\end{equation}
	 From the definition of limit, there exists a positive number 
	 \begin{equation}
	 \varepsilon \triangleq \min\big\{z_{(k-1)}-z_{(k)},z_{(k)}-z_{(k+1)}\big\},
	 \end{equation}
	 and a sufficiently large number $M$ such that 
\begin{equation}
	 \big|w_i(t)-z_{(k)}\big|<\varepsilon,\ \ t\ge M.
\end{equation}
	 This implies that after $M$ rounds of local communication, the thresholds $w_i(t)$ will lie in $(z_{(k+1)},z_{(k-1)})$ for all $i \in \{1,\cdots,n\}$. Furthermore, for $t\ge M$, the number of transmissions will be either $k$ or $k-1$. Therefore, either the $k$ or $k-1$ largest measurements will be sent to the remote estimator, resulting in the following inequality:
		\begin{equation}
			\frac{1}{n}\sum_{i=k+1}^{n} z_{(i)}^2 \le \mathcal{J}_Q\big(\vx,\vu(t) \big)\le \frac{1}{n}\sum_{i=k}^{n} z_{(i)}^2.	
		\end{equation}
	\end{IEEEproof}

	

\begin{figure}[!t]
		\centering
		\includegraphics[width=0.6\linewidth]{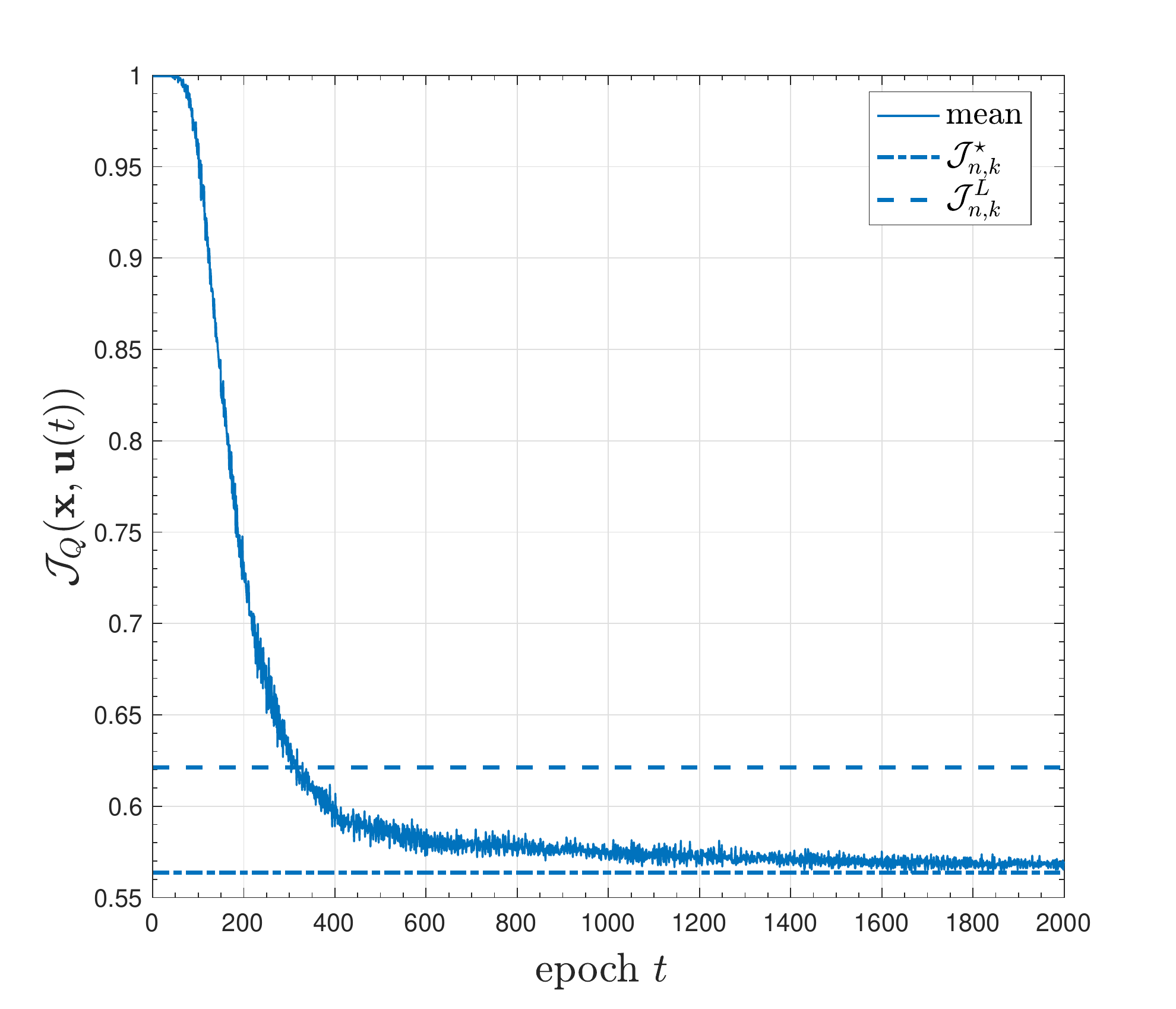}
		\caption{Mean of $500$ sample paths $\mathcal{J}_Q(\boldsymbol{x},\boldsymbol{u}(t))$. The dashed horizontal line represents the performance of the optimal decentralized scheme $\mathcal{J}_{n,k}^\star$. The dashed and dotted horizontal line represents the lower bound $\mathcal{J}_{n,k}^L$.}
		\label{fig: DecentralizedJvst_ErdoRenyi}
	\end{figure}

	 	\Cref{fig: DecentralizedJvst_ErdoRenyi} illustrates the performance of the scheme based on quantile estimation by computing the mean of $500$ sample paths $\mathcal{J}_Q(\boldsymbol{x},\boldsymbol{u}(t))$. The underlying graph is the same used in the simulation results in \cref{sec: consensus} and the observations are standard Gaussian random variables. The step-size sequence in \cref{eq:step} is defined with $\alpha=1000$, and $\tau=0.51$.
	 	
	 	 Comparing \cref{fig: performance_consensus.pdf,fig: DecentralizedJvst_ErdoRenyi}, we notice that the asymptotic performance of the quantile estimation scheme is superior to the performance of the consensus-based decentralized scheme, where the variance is first estimated followed by threshold computation. However, the convergence rate of the quantile estimation scheme is considerably slower than the consensus-based scheme. We provide a simple intuitive argument for these performance differences: The asymptotic performance of the quantile estimation scheme is essentially the performance of the optimal centralized scheme, however, in order to achieve it, the sensors must exchange much more information than what is needed to estimate the variance of the distribution via average consensus. Moreover, the quantile estimation scheme eliminates the occurrence of collisions in the long run, which cannot be avoided via the consensus-based scheme. Hence, the existence of a gap in performance.

	\subsection{Fast quantile estimation decentralized scheme}
	
	In this section, we introduce a hybrid scheme with a faster convergence rate and better or equal performance than both schemes presented so far.
	Let $R$ be an integer such that when $t < R$, we use the consensus-based method, which has a faster convergence rate; when $t=R$, each node uses the threshold computed by solving the optimization problem in \cref{eq:local_threshold} to initialize the quantile estimation scheme, i.e.,
	\begin{equation}\label{eq:quantile_init}
	w_i(R) = T_i^\star(R), \ \ i \in\{1,\cdots,n\}.	\end{equation}
    After that, we use the quantile estimation scheme as described in \cref{sec:quantile-based}. which converges to a smaller asymptotic cost. The instantaneous cost function $\mathcal{J}_F\big(\boldsymbol{x},\boldsymbol{u}(t)\big)$ is defined as follows:
\begin{equation}	
\mathcal{J}_F\big(\boldsymbol{x},\boldsymbol{u}(t)\big)=
	\left\{\begin{array}{ll}{\mathcal{J}_C\big(\boldsymbol{x},\boldsymbol{u}(t)\big)}, \quad {\text { if }  t<R,} \\ {\mathcal{J}_Q\big(\boldsymbol{x},\boldsymbol{u}(t)\big)}, \quad {\text { if } t\ge R.}\end{array}\right.
\end{equation}

    The switching time $R$ is chosen as the time when the local estimates in the consensus scheme approaches the average of the all of the square of the measurements. From that point on the decentralized system starts to behave as centralized.
	The convergence rate of the consensus-based scheme is determined by the matrix $(\mathbf{I}-{d_{\max}^{-1}}  \mathbf{L})$, where $\mathbf{L}$ is the graph Laplacian matrix. It is a well known fact \cite{xiao2004fast} that: 
\begin{equation}
\lim_{t \to +\infty} \big(\mathbf{I}-\frac{1}{d_{\max}}  \mathbf{L}\big)^t= \frac{1}{n}\mathbf{1 1}^{T}.
\end{equation} 
Therefore, we set $R$ to be
	\begin{equation}
	R(\delta)=\min \Bigg\{ t \ \Bigg| \  \norm{ \big(\mathbf{I}- \frac{1}{d_{\max}}\mathbf{L}\big)^{t}-\frac{1}{n} \mathbf{1 1}^{T}}\le \delta \Bigg\},
	\end{equation}
where $\delta \in (0,1)$ is a design parameter that can be chosen by cross-validation analysis performing  simulations. The resulting switching time $R$ is a function of the properties of the local connectivity graph $\mathbb{G}$. For the graph that we have been using for our numerical results, \cref{tab:switch} contains a few instances of switching time $R$ corresponding to different values of $\delta$. The underlying graph $\mathbb{G}$ is a sample from the Erdos-Renyi ensemble of random graphs with $n=1000$ nodes, connected by edge with probability $p_{\text{edge}}=0.05$. The graph $\mathbb{G}$ has $d_{\max}=81$ and $\lambda_2=27.35$ and is the same used in all numerical examples in this paper.

\begin{table}[h!]
    \centering
    \caption{Switching time $R$ for different values of $\delta$}
    \begin{tabular}{cc}
      $\delta$   & $R$ \\
     \hline \hline
       $1$  & 1 \\
       $0.1$  & 6\\
       $10^{-2}$  & 12 \\
       $10^{-3}$  & 17\\
       $10^{-4}$  & 23\\
       $10^{-5}$  & 28\\
       \hline \hline
    \end{tabular}
    \label{tab:switch}
    \vspace*{-0.1in}
\end{table}

	\subsection{Mistmatched distributions and illustrative example} \label{sec: Simulations}
	
	In order to broaden the applicability of our schemes, consider the case when the distribution is not necessarily Gaussian, but we use a Gaussian approximation, i.e., we perform the local threshold design assuming that the measurements are drawn from a Gaussian distribution with unknown variance. In this ``mismatched'' distribution scenario, we use consensus to estimate the variance of the distribution. At each iteration, each sensor solves \cref{eq: optimality} assuming the distribution is $\mathcal{N}\big(0,y_i(t)\big)$. At a certain point $t=R$, we initialize the quantile estimation scheme using \cref{eq:quantile_init}. 

\begin{figure}[t]
		\centering
\includegraphics[width=0.6\linewidth]{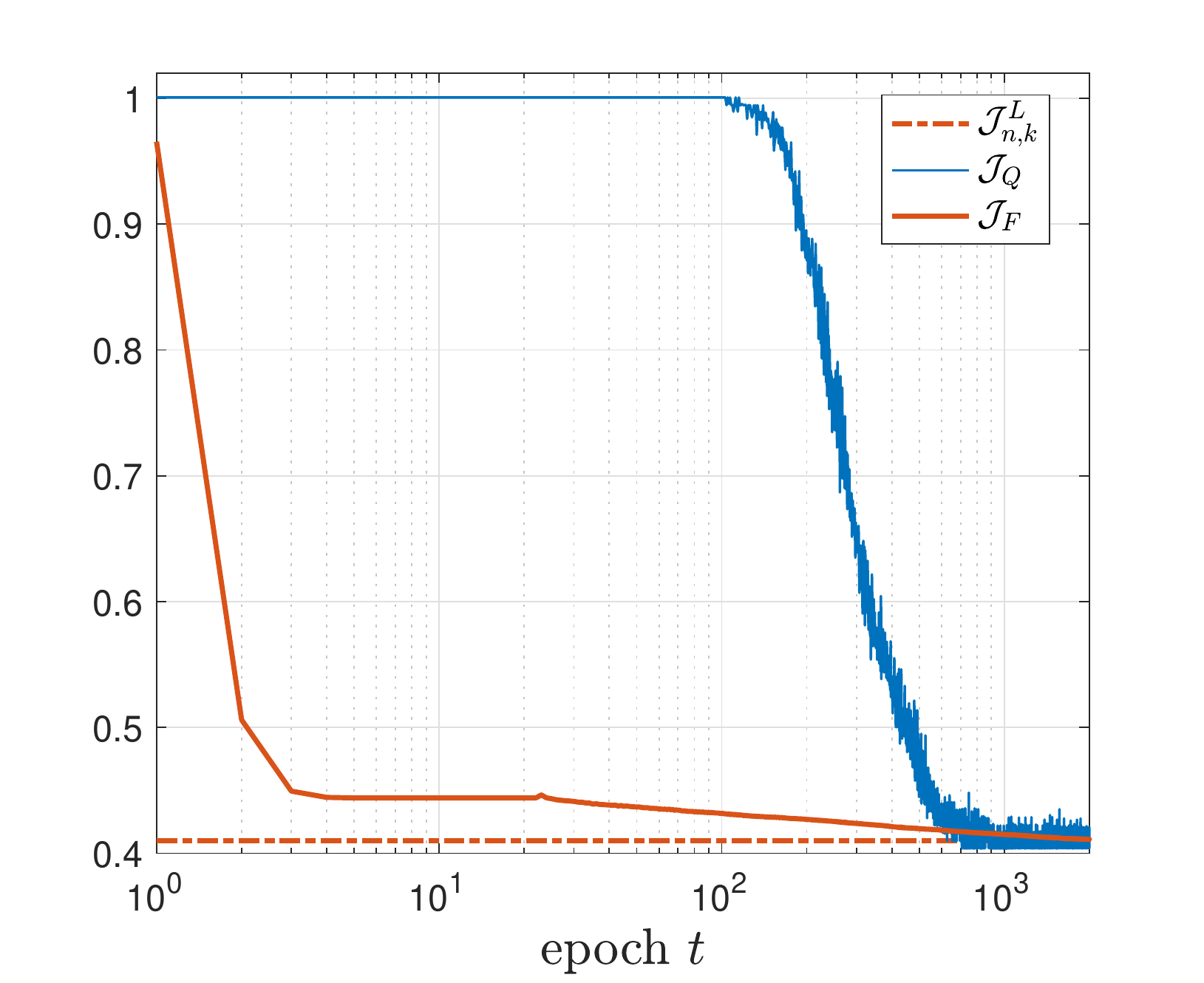}
		\caption{Performance of the quantile estimation scheme and its accelerated counterpart. The observations are i.i.d. according to a Laplacian distribution}
		\label{fig: Mismached Case}
	\end{figure}

The parameter $\delta$ used in the Fast quantile estimation scheme is chosen to be $10^{-4}$, which implies in a switching time $R=23$. Assuming the measurements are drawn from a Laplacian distribution, $X \sim \mathcal{L}(0,1)$, we compare the performance of the regular quantile estimation scheme with its accelerated counterpart. The numerical results shown in \cref{fig: Mismached Case} show that the fast quantile estimation scheme is approximately 2 orders of magnitude faster, even when the design is done based on mismatched distributions. The reason why the hybrid scheme is so effective, is that the consensus scheme quickly synchronizes the local estimates at the sensors and then all the sensors adjust the same threshold that will eventually match the $k$-th ordered statistics. This synchronization also leads to smooth sample paths, whereas the sample paths of the pure quantile estimation scheme display large oscillations when different local estimates are approaching the $k$-th ordered statistics. Finally, since that quantile estimation scheme is independent of the distributions, the asymptotic performance is unaffected by the mismatch. That is the reason why the hybrid scheme is able to achieve the desirable features of both schemes.

	\section{Conclusions}\label{sec: Conclusions}
	
	We have studied the design of a threshold strategy for a decentralized, remote estimation system over the collision channel with and without local communication. The channel allows at most $k\leq n$ simultaneous transmissions. Assuming that there is no local communication among the sensors and the distribution of the measurements is symmetric, our theoretical analysis shows the existence and uniqueness of an optimal threshold under the normalized MSE criterion.
    
    Our numerical results show that the decentralized scheme based on the optimal threshold strategy has a performance reasonably close to the optimal centralized scheme. Assuming there is local communication among sensors and the knowledge of the distribution of the observations is incomplete, three decentralized schemes have been proposed: the consensus-based, quantile estimation, and the fast quantile estimation schemes.  The consensus-based scheme has a faster convergence rate, while the quantile estimation scheme is the better asymptotic performance. By combining the two schemes, a fast quantile estimation scheme is derived, which combines the positive aspects of both strategies. 
    
    Several possible research directions can be pursued in the future. One is to introduce a penalty for delay in the performance metric, and determine what the optimal switching time that characterizes the fast quantile estimation scheme is. A second is to study the problem of learning through local interactions when the probabilistic model is entirely unknown. Finally, study the issue of privacy in the remote estimation, when the sensors would like to agree on a common strategy but are constrained by not disclosing their private information to neighboring nodes.

	\appendices
	
	\section{Proof of Lemma \ref{lm: cost function2}} \label{appendix: cost function2}
	We begin by defining the following event: 
	\begin{equation}
	\mathfrak{E}_{i,k}\triangleq \Big\{\sum_{\ell \neq i} D_\ell \le k-1\Big\} \end{equation}
	and its complement: 
	\begin{equation} 
	\mathfrak{E}^c_{i,k} \triangleq \Big\{\sum_{\ell \neq i} D_\ell \ge k\Big\}.
	\end{equation}
	
	Using the law of total expectation, the objective function in Eq.~(\ref{eq: MSE}) becomes
	\begin{align}
	&\mathcal{J}_{n,k}(T)=\frac{1}{n}\sum_{i=1}^{n}\Bigg[\E\[\(X_i-\hat{X}_i\)^2\ \Big{|}\ D_i=0\] \P(D_i=0) \nonumber\\
	&+\E\[\(X_i-\hat{X}_i\)^2\ \Big{|}\  D_i=1, |\mathbb{D}|> k\] \P(D_i=1,|\mathbb{D}|> k)
	\nonumber\\
	&+\E\[\(X_i-\hat{X}_i\)^2\ \Big{|}\  D_i=1, |\mathbb{D}|\le k\] \P(D_i=1,|\mathbb{D}|\le k)\Bigg].
	\end{align}
	
	Applying the independence of $D_i$ and $\{D_\ell\}_{\ell \neq i}$ yields
	\begin{align}\label{eq:cost_split}
	&\mathcal{J}_{n,k}(T)=\frac{1}{n}\sum_{i=1}^{n}\Bigg[\E\[\(X_i-\hat{X}_i\)^2\ \Big{|}\ D_i=0\] \P(D_i=0)\nonumber\\
	&+\E\[\(X_i-\hat{X}_i\)^2\ \Big{|}\  D_i=1, \mathfrak{E}^c_{i,k}\] \P(D_i=1)\P(\mathfrak{E}^c_{i,k})\nonumber\\
	&+\E\[\(X_i-\hat{X}_i\)^2\ \Big{|}\  D_i=1, \mathfrak{E}_{i,k}\] \P(D_i=1)\P(\mathfrak{E}_{i,k})\Bigg].
	\end{align}
	
	Substituting Eq.~(\ref{eq: estimator1}) in Eq.~(\ref{eq:cost_split}) yields
	\begin{equation}
	\mathcal{J}_{n,k}(T)=\frac{1}{n}\sum_{i=1}^{n} \Bigg[ \E\[X_i^2 \ \Big{|}\ D_i=0\] \P(D_i=0)  \\
	+ \E\[ X_i^2 \ \Big{|} \ D_i=1\] \P(D_i=1)\P(\mathfrak{E}^c_{i,k}) \Bigg].
	\end{equation}
	
	Therefore, the cost function becomes
	\begin{align}
	\mathcal{J}_{n,k}(T) & = \frac{1}{n} \sum_{i=1}^n  \bigg[ \E\[X_i^2  \mathbf{1}\({D_i=0}\)\] + \E\[X_i^2   \mathbf{1}\({D_i=1}\)\] \P(\mathfrak{E}^c_{i,k}) \bigg] \\
	& = \frac{1}{n} \sum_{i=1}^n  \bigg[ \E\[X_i^2\]  - \E\[X_i^2   \mathbf{1}\({D_i=1}\)\] \P(\mathfrak{E}_{i,k}) \bigg]. 
	\end{align}
	Since $\{X_i\}_{i=1}^n$ are i.i.d., after some elementary algebra we have:  
	\begin{equation} 
	\mathcal{J}_{n,k}(T) = \E\[X^2\]- \E\[X^2 \mathbf{1}(|X| \ge  T) \]F_{n,k}(T).
	\end{equation}
	
	\section{Proof of Theorem \ref{thm: optimality}} \label{appendix: optimality}
	
	
\begin{IEEEproof}[Proof of Theorem 1]
According to \cite{arnold1992first}, the derivative of $F_{n,k}(T)$   is
	\begin{equation} \label{OrderStatistics_pdf}
	f_{n,k}(T)=  2k \binom{n-1}{k}  p(T)^{n-1-k}\big(1-p(T)\big)^{k-1}f_X(T).
	\end{equation}
	
	
	We shall show that there is a unique $T^\star$ such that the derivative $\mathcal{J}_{n,k}'(T)$ is zero for $T>0$, where
	\begin{equation}
	\mathcal{J}_{n,k}'(T)=2T^2 f_X(T) F_{n,k}(T)-\E\[X^2 \mathbf{1}(|X| \ge T)\]f_{n,k}(T).
	\end{equation}
	
	Due to the fact that $f_X(T)>0$ for $T >0$, \cref{OrderStatistics_pdf} implies that  $f_{n,k}(T)>0$ for $T>0$. So we have
	\begin{equation} 
	\frac{\mathcal{J}_{n,k}'(T)}{f_{n,k}(T)}
		= \underbrace{\[ \frac{2T^2 f_X(T) F_{n,k}(T)}{f_{n,k}(T)}-\E\[X^2 \mathbf{1}(|X| \ge T)\]\]}_{ \triangleq h(T)}.
		\end{equation}
	Incorporating \cref{OrderStatistics_cdf} and  \cref{OrderStatistics_pdf} into $h(T)$ yields
	\begin{equation} \label{eq: h_T}
		h(T)=T^2 \cdot \frac{ \sum\limits_{j=0}^{k-1} \binom{n-1}{j}p(T)^{n-1-j}(1-p(T))^{j}}{k\binom{n-1}{k} p(T)^{n-1-k}(1-p(T))^{k-1}} 
		  -\E\[X^2 \mathbf{1}(|X| \ge T)\].
	\end{equation}
After some algebra, we obtain
\begin{equation}\label{eq: h_T_2}
h(T)=T^2p(T)\sum_{j=0}^{k-1} \frac{(k-1)!(n-1-k)!}{j!(n-1-j)!} 
		\(\frac{p(T)}{1-p(T)}\)^{k-j-1} -\E\[X^2 \mathbf{1}(|X| \ge T)\] .
\end{equation}
	Since $p(T) \in (0,1) $ is strictly increasing for $T >0$, $1/(1-p(T))\in (0,\infty)$ is also strictly increasing for $T>0$. The product of two positive and strictly increasing functions is a positive strictly increasing function, which implies that 
	$
	{p(T)}/(1-p(T))\in (0,+\infty)
	$
	is a strictly increasing function.
	
	Since $k-j-1\ge 0$ for $j \in \{0,\cdots,k-1\}$, we obtain that 
	\begin{equation}
	\[\frac{p(T)}{1-p(T)}\]^{k-j-1}\in (0,+\infty)
	\end{equation}
	is a non-decreasing function. The sum of non-decreasing functions is a non-decreasing function, which implies that
	\begin{equation}
	\sum_{j=0}^{k-1} \Bigg[ \frac{(k-1)!(n-1-k)!}{j!(n-1-j)!} \(\frac{p(T)}{1-p(T)}\)^{k-j-1}\Bigg]
	\end{equation}
	is a positive non-decreasing function.
	
	Using the fact that the product of a positive strictly increasing and a positive non-decreasing function is a strictly increasing function, we have
	\begin{equation}
	T^2  p(T)\sum_{j=0}^{k-1} \Bigg[ \frac{(k-1)!(n-1-k)!}{j!(n-1-j)!} \(\frac{p(T)}{1-p(T)}\)^{k-j-1}\Bigg]
	\end{equation}
	is a positive strictly increasing function.
	
	Combining with the fact that $-\E\[X^2 \mathbf{1}(|X| \ge T)\]$ is  strictly increasing continuous function for $T >0$, we conclude $h(T)$ is a strictly increasing continuous function of $T$ for $T >0$.
	
	When $T \to 0^{+}$, we have $\E\[X^2 \mathbf{1}(|X| \ge T)\]>0$ and $p(T)\to 0$, which implies
	\begin{equation}
	\inf_T h(T)= \lim_{T \to 0^{+}}h(T)<0.
	\end{equation}
	
	When $T \to +\infty$, we have $\E\[X^2 \mathbf{1}(|X| \ge T)\] \to 0$ and $p(T)\to 1$, which implies
	\begin{equation}
	\sup_T h(T)=\lim_{T \to +\infty}h(T)>0.
	\end{equation}
	
	Therefore, there exists only one $T^\star \in (0, +\infty)$ such that $h(T^\star)=0$. Since $f_{n,k}(T)>0$ and $\mathcal{J}_{n,k}'(T) = f_{n,k}(T) h(T)$, there is a unique $T^\star$ that minimizes $\mathcal{J}_{n,k}(T)$ for $T \in (0, +\infty)$.  Combining with the fact that $\mathcal{J}_{n,k}(T)$ is a continuous function,  $T^\star$ is the optimal threshold.

	\end{IEEEproof}

	\section{Proof of Lemma \ref{lm: initialization}} \label{Appendix: Proof_Lemma_initialization}
	Recalling that $h(T)$ in \cref{eq: h_T} is a strictly increasing continuous function of $T$, establishing the inequality 
	$
	T^\star > T_c
	$
	is equivalent to showing that
	\begin{equation}
	h\(T_c\) < h(T^\star)=0,
	\end{equation}
	where 
	\begin{equation}
	    T_c \triangleq p^{-1}\(1-\frac{k}{n}\).
	\end{equation}
	Using \cref{eq: h_T_2} yields
	\begin{IEEEeqnarray}{rCl}
	h\(T_c\)&=& T_c^2  \frac{n-k}{n}  \sum_{j=0}^{k-1}\frac{(k-1)!(n-1-k)!}{j!(n-1-j)!} \(\frac{n-k}{k}\)^{k-j-1}  -\E\[X^2 \mathbf{1}(|X| \ge T_c)\]  \\
	&=& \frac{T_c^2}{n}  \sum_{j=0}^{k-1}\frac{k!(n-1-k)!}{j!(n-1-j)!} \(\frac{n-k}{k}\)^{k-j}-\E\[X^2 \mathbf{1}(|X| \ge T_c)\] \\
	&<& \frac{T_c^2}{n}   \sum_{j=0}^{k-1}\frac{k!(n-1-k)!}{j!(n-1-j)!} \(\frac{n-k}{k}\)^{k-j}-\frac{kT_c^2 }{n}  \\
	&=&\frac{T_c^2}{n} \(\sum_{j=0}^{k-1}\frac{k!(n-1-k)!}{j!(n-1-j)!} \(\frac{n-k}{k}\)^{k-j}-k\) \\
	&\le& 0,
	\end{IEEEeqnarray}
	where the first inequality follows from
	\begin{IEEEeqnarray}{rCl}
	\E\[X^2 \mathbf{1}(|X| \ge T_c)\]&=&\int_{T_c}^{+\infty} x^2f_Z(x)dx  \\ 
	&>&T_c^2 (1-F_Z(T_c)) = \frac{kT_c^2}{n},
	\end{IEEEeqnarray}
	and the last inequality follows from
	\begin{IEEEeqnarray}{rCl}
	\sum_{j=0}^{k-1}\frac{k!(n-1-k)!}{j!(n-1-j)!} \(\frac{n-k}{k}\)^{k-j}
	&=&\sum_{j=0}^{k-1}\frac{k(k-1)\cdots(j+1)}{(n-j-1)\cdots(n-k)} \(\frac{n-k}{k}\)^{k-j} \\
	&=&\sum_{j=0}^{k-1}\frac{k(k-1)\cdots(j+1)}{k^{k-j}}
	\frac{(n-k)^{k-j}}{(n-j-1)\cdots(n-k)}  \\
	&\le& k.
	\end{IEEEeqnarray}

		\section{Proof of Lemma \ref{lm: phasetransition}} \label{Appendix: phasetransition}
	
	\begin{lemma}[Chernoff's inequality, \cite{mitzenmacher2017probability}] \label{prop: Chernoff's inequality}
		Let $X_1,\ldots,X_n$ be i.i.d. Bernoulli random variables with probability $p_i$. Let  $S_n=\sum_{i=1}^{n}X_i$ and $\mu=\E [S_n]$. Then 
		\begin{equation} \label{lm: chernoff_ge}
		\P\big( S_n \ge (1+\delta) \mu \big) \le \exp\(-\frac{\delta^2}{2+\delta} \mu \). \ \ \delta >0 ,
		\end{equation}
	\end{lemma}
	
	\begin{IEEEproof}[Proof of Lemma \ref{lm: phasetransition}]
	Since $\{D_i\}$ is an i.i.d. sequence of Bernoulli random variables with probability 
	\begin{equation}
	\P\(D_i=1\)=1-p(T),
	\end{equation}
	we have
	\begin{equation}
	\mu=\E \[\sum_{i=1}^{n-1} D_i \]=(n-1)\big(1-p(T)\big).
	\end{equation}
	
	
Let $\delta = \frac{k-\mu}{\mu}$. Then, the following holds
	\begin{IEEEeqnarray}{rCl}
	F_{n,k}(T) & = & 1-\P\(\sum_{i=1}^{n-1} D_i \ge k\) \\
		& \ge & 1- \exp \(-\frac{\(\mu-k\)^2}{k+\mu}\) \\
	&\ge &  1- \exp \(-\frac{\(\mu-k\)^2}{2k}\).
	\end{IEEEeqnarray}
Define
\begin{equation}
T_r(s) \triangleq p^{-1}\(1-\frac{k-s\sqrt{2k}}{n-1 }\).
\end{equation}
Therefore,
\begin{equation}
F_{n,k}\big(T_r(s)\big)\geq 1 - \exp(-s^2).
\end{equation}

\end{IEEEproof}

\bibliographystyle{IEEEtran}
	\bibliography{IEEEabrv,refs}	
	
	\end{document}